\documentclass[twocolumn,showpacs,preprintnumbers,amsmath,amssymb,aps,prb]{revtex4-2}

\usepackage{graphicx}
\usepackage{booktabs}
\usepackage{amsmath}
\usepackage{amssymb}
\usepackage{hyperref}
\usepackage{xcolor}
\usepackage{multirow}
\usepackage{tikz}
\usetikzlibrary{calc,positioning,fit,backgrounds,decorations.pathreplacing,arrows.meta}

\newcommand{\CLOPSh}{\textsc{CLOPS}\ensuremath{_h}}
\newcommand{\CLOPSv}{\textsc{CLOPS}\ensuremath{_v}}
\newcommand{\CLOPS}{\textsc{CLOPS}}
\newcommand{\QV}{\textsc{QV}}
\newcommand{\LF}{\textsc{LF}}
\newcommand{\LFOC}{\textsc{LFOC}}
\newcommand{\EPLG}{\textsc{EPLG}}
\newcommand{\CPS}{\textsc{MCPS}}

\begin{document}

\title{CLOPS: Benchmarking System Speed at Utility Scale}

\author{Andrew Wack}
\affiliation{IBM Quantum, IBM T.\ J.\ Watson Research Center, Yorktown Heights, NY 10598, USA}


\date{\today}

\begin{abstract}
As quantum processors scale to hundreds of qubits, execution speed is a
critical performance dimension alongside scale and quality. While substantial
progress has been made in benchmarking circuit fidelity, existing speed metrics
often fail to reflect the sustained, end-to-end throughput experienced by users
running utility-scale workloads. This shortfall is especially pronounced for
layered, parameterized circuits executed repeatedly within classical-quantum
workflows, such as variational algorithms and error-mitigated simulations.

In this work, we formalize \CLOPSh{} (Circuit Layer Operations Per Second) as
a holistic speed benchmark defined over layered, hardware-aware circuits.
\CLOPSh{} measures the sustained rate at which the system executes
\emph{physical layers}---parallel slices of qubit-disjoint two-qubit gates
separated by synchronization barriers. Because each such layer is one time slice
of an $N$-qubit circuit, this rate maps directly to the execution rate of
layered $N$-qubit circuits, connecting \CLOPSh{} to published device capability
claims, and to the device-level throughput ceiling we formalize as Max Circuits
Per Second (\CPS{}). \CLOPSh{} is
obtained under layer-fidelity operating conditions,
binding the speed measurement to an independently verified quality envelope, and
it shares its layer decomposition with scalable layer-fidelity (\LF{}) quality
benchmarks---enabling coherent interpretation of speed and quality without
conflating the two.
\end{abstract}

\maketitle

\section{Introduction}

\subsection{Performance Metrics in the Utility Era}

The performance of a quantum computing system is governed by three
interdependent attributes: scale, quality, and speed. Scale determines the size
of problems that can be represented; quality determines the depth and structure
of circuits that can be executed reliably; and speed determines how much useful
work can be completed per unit time. This framing has been articulated
previously as a foundation for benchmarking near-term quantum computers and
remains increasingly relevant as systems enter the regime of utility-scale
experimentation~\cite{wack2021}.

Recent progress in hardware and software has enabled the execution of quantum
circuits spanning 100 or more qubits. At the same time, the dominant workloads
of practical interest have shifted toward iterative, layered algorithms embedded
within classical optimization or post-processing loops, including variational
methods, Trotterized dynamics, and error-mitigated
experiments~\cite{kandala2017,farhi2014,temme2017,kim2023}. In
such settings, wall-clock runtime is often dominated not by gate durations but
by the sustained throughput at which these circuits can be executed repeatedly.

In this context, execution speed becomes a first-order constraint. Even when a
circuit can be executed with acceptable fidelity, the overall wall-clock time
required to generate statistically meaningful results may render an experiment
impractical. As error mitigation techniques trade circuit repetition for
improved accuracy, the total computational cost scales rapidly, amplifying the
importance of sustained execution
throughput~\cite{temme2017,endo2018,mckay2023}. Consequently,
meaningful progress toward quantum advantage requires not only improvements in
fidelity and scale, but also systematic gains in the speed at which quantum
experiments can be executed.

\subsection{Why Speed Must Be Benchmarked at the Right Level}

Benchmarking speed in quantum computing systems requires distinguishing between
performance at different levels of abstraction. Quantum systems---like classical
computing systems before them---support a hierarchy of benchmarks, each valid
and informative at its own level, but each addressing a different set of
questions. Table~\ref{tab:hierarchy} organizes this hierarchy across three
performance dimensions---scale, quality, and speed---at four levels of
abstraction.

\begin{table*}[t]
\caption{\label{tab:hierarchy}Representative benchmarks organized by abstraction
level and performance dimension. Entries are examples only; the table is not
exhaustive. Empty cells indicate no widely-adopted metric currently exists for
that combination. Cliffordized proxy circuits are the scalable, verifiable lift
of the device-level layer-fidelity/\EPLG{} entry to full circuits: full-circuit
fidelity is recovered from layer fidelities by fidelity multiplication and
verified at scale using Clifford proxies~\cite{merkel2025}.}
\begin{ruledtabular}
\begin{tabular}{lccc}
\textbf{Level} & \textbf{Scale} & \textbf{Quality} & \textbf{Speed} \\
\hline
Qubit  & Physical qubits per programmable qubit & Gate fidelity, $T_1/T_2$ & Gate time \\
Device & Programmable qubit count & Layer Fidelity / \EPLG{} & \CPS{} \\
System & (TBD) & \QV{}, Cliffordized proxy circuits & \CLOPSh{} \\
Application & (TBD) & (TBD) & (TBD) \\
\end{tabular}
\end{ruledtabular}
\end{table*}

Benchmarks at each level are meaningful on their own terms. Qubit-level metrics
expose fabrication quality and isolate component performance under ideal
conditions. Even scale is non-trivial at this level: the programmable qubit
exposed to an algorithm is not equivalent to a single physical qubit, and
architectures differ in how many physical qubits are required to realize one
programmable qubit~\cite{ibmhardware2026}. Device-level metrics characterize the processor at scale, e.g. what
fidelity the full coupling graph delivers under realistic parallel operation, 
the maximum number of circuits per second (\CPS{}; defined in
Sec.~\ref{sec:mcps}) under the same operating conditions, etc. Lower-level benchmarks also provide natural guard rails on what
higher-level benchmarks can achieve: the \CPS{} of the device sets a hard
ceiling on system-level throughput that no software optimization can exceed.

The critical issue arises when metrics from one level are used to predict
performance at a higher level. Gate times do not strictly predict system throughput;
device repetition rates do not predict the execution rate of full parameterized
circuits. The gaps between levels are filled by software compilation, parameter
binding, classical co-processing, data movement, and control-system
overhead---none of which appear in lower-level measurements. Classical
high-performance computing encountered the same problem: peak floating-point
rates at the chip level consistently overestimated achievable system throughput,
leading to the adoption of system-level benchmarks such as LINPACK, which
measure sustained performance on a representative workload across the full
memory and interconnect hierarchy~\cite{dongarra2003}. The analogous lesson
applies to quantum systems: characterizing execution speed accurately requires a
benchmark defined at the system level, one that is inclusive of all classical
and quantum costs associated with a complete execution pipeline~\cite{wack2021}.

A further complication arises from the distinction between latency and
throughput. Short-duration benchmarks that measure the execution time of a small
number of circuits are often dominated by transient effects such as compilation
warm-up or control loading~\cite{wack2021}. Many practical workloads operate in a steady-state
regime, where pipelines are filled and circuits are executed continuously over
extended periods. Benchmarks that do not explicitly target steady-state behavior
risk mischaracterizing system performance for long-running experiments.

The application level of this hierarchy---benchmarks tied to specific
computationally relevant problems, analogous to the SPEC suite in classical
computing~\cite{henning2006}---remains largely undeveloped for quantum systems. Meaningful
application benchmarks require workloads with established scientific or business
value, known classical reference points, and problem sizes that are beyond
efficient classical simulation---otherwise the benchmark cannot distinguish
quantum execution from a classical solver.

Several efforts have approached this level. The QED-C application-oriented
benchmark suite~\cite{lubinski2023} measures result fidelity and
quantum execution time across a range of algorithm families, and represent the
most systematic attempt to date to benchmark hardware at the algorithm level.
However, they operate on circuits well below utility scale---within the
classically simulatable regime---do not report classical compilation or
optimization overhead, and measure single-execution quality rather than sustained
throughput under load, falling short of the time-to-solution metric their
framing implies. The Metriq platform~\cite{cosentino2026} aggregates system- and
application-inspired metrics into a composite score but explicitly does not
impose a benchmark hierarchy, covers circuits well below utility scale, and
excludes end-to-end classical overhead. Neither framework constitutes an
application-level speed benchmark in the sense of measuring how rapidly a system
delivers correct, useful output at scale under realistic workload conditions.
That said, applications are beginning to reach the necessary regime: sample-based
Krylov quantum diagonalization (SKQD) has been applied to model protein-ligand
complexes exceeding 12{,}000 atoms~\cite{merz2026}, the kind of domain-relevant,
beyond-classical workload that could anchor a future application-level benchmark.

For now, the system level is the highest level at which principled,
hardware-agnostic benchmarking is currently tractable. These considerations
motivate the need for a system-level speed benchmark that
is (i)~defined in terms of parallel circuit layers, (ii)~inclusive of all
classical and quantum costs associated with execution, (iii)~measured in
steady-state operation, and (iv)~executed under operating conditions anchored to
an independently verified quality envelope.

\subsection{From Abstract Speed Metrics to Layer-Aligned Benchmarks}

Early attempts to benchmark execution speed, including the original definition
of Circuit Layer Operations Per Second (CLOPS, referred to here as
\CLOPSv{}), addressed this need by measuring how quickly a system could execute
parameterized quantum circuits while interacting with the runtime
environment~\cite{wack2021}. \CLOPSv{} leveraged circuit families derived from
Quantum Volume (QV) experiments, ensuring a connection to a well-established quality
metric.

In this paper, we re-instantiate the \CLOPS{} framework as \CLOPSh{} by
defining speed over explicitly layered, hardware-aware circuits whose structure
is directly compatible with \LF{} benchmarking. This alignment enables coherent
interpretation of execution throughput in the context of utility-scale systems,
while preserving the holistic, implementation-agnostic character that motivated
the original proposal.

\section{Background and Related Work}

\subsection{Quality Benchmarks for Large-Scale Quantum Devices}

Benchmarking the quality of quantum processors has been an active area of
research since the earliest experimental demonstrations of programmable quantum
systems. Traditional approaches focus on characterizing individual
components---single- and two-qubit gate fidelities, measurement errors, and
coherence times---using protocols such as randomized
benchmarking~\cite{knill2008,magesan2012}, but these component-level metrics
are insufficient to characterize system performance as experienced by users
executing nontrivial algorithms~\cite{wack2021}.

Holistic benchmarks based on executing structured or randomized circuits address
this gap. Among these, Quantum Volume (\QV{}) has emerged as a widely adopted
metric that simultaneously incorporates gate fidelity, connectivity, compilation
efficiency, and hardware control performance~\cite{cross2019}. \QV{} measures
the largest square circuit of random $\mathrm{SU}(4)$ layers that a system can
execute with sufficiently high heavy-output probability. Despite its success,
\QV{} has important limitations at utility scale: it reports on the
best-performing subset of qubits rather than whole-device performance, its
pass-fail nature makes it insensitive to incremental improvements, and its
unstructured random circuits differ substantially from the repetitive layered
structures used in many near-term algorithms~\cite{cross2019,mckay2023}.

To address these limitations, layer fidelity (\LF{}) has been introduced as a
scalable quality benchmark for large quantum processors~\cite{mckay2023}. \LF{}
measures the fidelity of executing a full layer of two-qubit
gates across a connected region of the device, capturing crosstalk and
idle-qubit errors. The associated quantities---error per layered gate (\EPLG{})
and the mitigation parameter $\gamma$---enable direct reasoning about
algorithmic runtime and error-mitigation cost. \LF{} results can, in turn, be
translated back into full-circuit fidelities by fidelity multiplication, and
verified directly and scalably using Cliffordized-circuit
approaches~\cite{merkel2025}. Importantly, an \EPLG{} layer
covers all interactions in the layer, and is therefore realized as a
fixed sequence of qubit-disjoint sublayers executed in order. This composite 
structure is the \LF{} notion of a ``layer'': an ordered
sequence of physical sublayers, each of which is a single parallel execution
slice of qubit-disjoint two-qubit gates. These physical sublayers are the same
execution units that \CLOPSh{} uses as its counting unit
(Fig.~\ref{fig:decomposition}).

\subsection{\label{sec:mcps}Speed Metrics at the Device and System Levels}

Principled speed metrics for quantum systems have lagged behind quality
benchmarks. To understand why a system-level speed benchmark is needed, it is
useful to examine what existing speed indicators capture and where their scope
ends.

Gate-per-second metrics characterize the raw throughput of the quantum processor
itself---how quickly the hardware can apply primitive operations under laboratory
conditions. These are meaningful qubit-level measurements: they expose such things as
fabrication and control advances, Hamiltonian design, etc., and a processor that 
cannot achieve competitive
gate rates cannot achieve competitive system throughput. Their limitation is
scope, not validity. Gate rates measure individual gates in isolation; they exclude
the parallel execution structure of real circuits, in which many gates execute
simultaneously within a layer, and they exclude the classical processing and
data-transfer overheads that impact end-to-end runtime at the system
level~\cite{wack2021}.

At the device level, a more comprehensive speed indicator accounts for the full
circuit cycle across the entire processor. \CPS{} captures this as the maximum
rate at which a device can execute a minimal-depth circuit repeatedly:
\begin{equation}
  \CPS{} = \frac{1}{t_{\mathrm{gate}} + t_{\mathrm{sp}} + t_{\mathrm{m}} + t_{\mathrm{delay}}},
  \label{eq:mcps}
\end{equation}
where $t_{\mathrm{gate}}$ is the duration of a Hadamard gate applied
simultaneously to all qubits on the device, $t_{\mathrm{sp}}$ is the 
state preparation time, $t_{\mathrm{m}}$ is the measurement time, 
and $t_{\mathrm{delay}}$ is the minimum
inter-circuit delay required to maintain calibration-consistent state
preparation and measurement fidelity. \CPS{} is a device-level ceiling
metric---it captures the fastest the hardware can cycle for a trivial
single-qubit gate circuit, and any circuit containing two-qubit gates will execute
strictly slower. This device-level speed
metric sets a hard ceiling on what any higher-level benchmark can achieve. If
the device cannot sustain a \CPS{} rate, the system cannot either. However,
\CPS{} is insensitive to circuit depth and other characteristics of the control
system. For utility-scale circuits involving many parallel layers and explicit
synchronization barriers, \CPS{} components are just a few of the many factors
that matter for algorithm runtime.


Finally, most device-level speed characterizations do not capture parameterized
execution, a defining feature of modern quantum workloads. Variational algorithms
and layered error mitigation require repeated execution of the same circuit
template with different parameter instantiations. The costs of parameter binding,
payload construction, and data movement are integral to total execution time at
the system level but invisible at the device level. Benchmarks based on fixed,
fully precompiled circuits systematically underestimate runtime for iterative
workloads~\cite{wack2021,murali2020}.

A separate and complementary axis is the benchmarking of the software stack
itself. Suites such as Benchpress~\cite{nation2025} characterize the
cost of \emph{producing} a hardware-executable circuit from an abstract
algorithm---transpilation, qubit routing and layout, gate synthesis, and
scheduling---across circuits and software packages. This is orthogonal to
execution speed: a system that constructs circuits quickly need not execute them
quickly, and vice versa. \CLOPSh{}, by contrast, measures the cost of
\emph{repeatedly executing} an already-prepared circuit template, not the cost
of preparing it. How
circuit construction cost does or does not enter a \CLOPSh{} measurement is made
precise in Section~\ref{sec:definition} through the steady-state execution
protocol.

In summary, gate and device level speed metrics are valid and informative within their
scope. The problem is not that they exist, but that system-level performance
cannot be inferred from them---the software stack, classical co-processing, and
execution pipeline between them and the user are opaque to device-level
measurement. A system-level benchmark is required to close this gap.



\section{\label{sec:evolution}Evolution of the \CLOPS{} Framework}

\subsection{The Original \texorpdfstring{\CLOPSv{}}{CLOPS\_v} Benchmark}

\CLOPSv{} was designed to measure how many circuit layers a quantum computing
system could execute per unit time while accounting explicitly for classical
interactions such as circuit submission, runtime compilation, parameter updates,
and result retrieval~\cite{wack2021}. It used parameterized Quantum Volume
circuit templates, fixing the random circuit structure offline and varying gate
parameters at runtime. This modeled the execution pattern of iterative
algorithms such as variational quantum eigensolvers and ensured that both quantum
execution and classical runtime costs were included.

\CLOPSv{} demonstrated that, for contemporary devices, classical overheads---
particularly runtime compilation, data transfer, and control-system
initialization---often dominated total execution time~\cite{wack2021}. These insights guided
improvements in runtime architectures and compilation pipelines.

\subsection{Limitations of \texorpdfstring{\CLOPSv{}}{CLOPS\_v}}

Despite these contributions, \CLOPSv{}'s reliance on \QV{}-derived circuits
introduced several shortcomings.

First, \QV{} circuits are inherently constrained by the largest classically
simulatable system sizes and are therefore limited in width and effective depth~\cite{cross2019}.
As quantum processors scaled to hundreds of qubits, these circuits no longer
reflected the sizes or connectivities of practical applications.

Second, \QV{} layer structure---random all-to-all $\mathrm{SU}(4)$
permutations---differs from the hardware-aware, synchronization-bounded layers
used in algorithmic workloads~\cite{cross2019}. This mismatch made it difficult to relate
measured speed to the performance of structured workloads.

Third, \QV{}'s sampling problem does not lend itself to the use of many error mitigation techniques.
As error-mitigated experiments have become a dominant paradigm at utility scale,
\CLOPSv{}'s circuit family became structurally incompatible with the workloads
users actually run.

\subsection{Design Goals of \texorpdfstring{\CLOPSh{}}{CLOPS\_h}}

The evolution from \CLOPSv{} to \CLOPSh{} was guided by five design goals:

\begin{enumerate}
\item \textit{Utility-scale representativeness.} The benchmark should operate on
  circuits comparable in width and depth to contemporary experiments, stressing
  the runtime, compiler, control electronics, and data paths at scale.

\item \textit{Layer-centric abstraction.} Speed should be expressed in terms of
  parallel circuit layers, in line with algorithmic complexity and hardware
  execution semantics.

\item \textit{Steady-state throughput.} Measurement should focus on sustained
  execution rate after pipelines are filled, excluding transient effects.

\item \textit{Implementation agnosticism.} The benchmark should permit a wide
  range of software and hardware optimizations, provided all relevant execution
  costs are included.

\item \textit{Quality anchoring.} \CLOPSh{} should be executed under operating
  conditions consistent with an independently measured quality benchmark,
  preventing speed-quality tradeoffs from inflating results.
\end{enumerate}

\subsection{Adoption of Layered, Hardware-Aware Circuits}

To satisfy these goals, \CLOPSh{} adopts a circuit family built from
hardware-aware layers rather than \QV{}-style random circuits. Each physical
layer consists of qubit-disjoint two-qubit gates arranged according to the
device's coupling graph, supplemented by fully twirled single-qubit operations
on all qubits including idling qubits. Global barriers are enforced at each
layer boundary, ensuring that all operations in a layer complete before the next
layer begins.

Importantly, \CLOPSh{} does not perform fidelity estimation. It executes
parameterized instances of a fixed layered circuit repeatedly, measuring how
quickly the system can sustain this execution pattern. Throughput is therefore
isolated as an independent performance dimension while maintaining structural
compatibility with quality benchmarks.

\begin{table*}[t]
\caption{\label{tab:evolution}Summary of changes from \CLOPSv{} to \CLOPSh{}.}
\begin{ruledtabular}
\begin{tabular}{p{2.8cm}p{6.5cm}p{6.5cm}}
\textbf{Property} & \textbf{\CLOPSv{}} & \textbf{\CLOPSh{}} \\
\hline
Circuit family   & QV-derived parameterized circuits                          & Layered, hardware-aware circuits with explicit synchronization barriers \\
Scale            & Small-to-medium circuits constrained by QV                 & Utility-scale ($\mathcal{O}(100)$ qubits and layers) \\
Layer semantics  & Implicit structure inherited from QV                       & Explicit physical layer derived from LF layers \\
Quality anchor   & Implicit (QV conditions)                                   & Explicit layer-fidelity operating conditions declared at reporting time \\
Benchmark role   & Exploratory system-level speed benchmark                   & Formalized speed benchmark for layered, parameterized execution \\
\end{tabular}
\end{ruledtabular}
\end{table*}

\subsection{\texorpdfstring{\CLOPSh{}}{CLOPS\_h} as a Re-instantiation of
the \texorpdfstring{\CLOPS{}}{CLOPS} Framework}

\CLOPSh{} is a re-instantiation of the \CLOPS{} framework for the utility-scale
regime---not a refinement of \CLOPSv{}, and not a replacement of the underlying
concept. It preserves the defining design principles of \CLOPSv{}: holistic
inclusion of the quantum-classical stack, sensitivity to runtime bottlenecks,
and flexibility in implementation. What changes is the circuit family and
quality anchor, chosen to match the regime in which the benchmark operates
(Table~\ref{tab:evolution}). By
grounding the benchmark in hardware-aware layered circuits and explicit
layer-fidelity operating conditions, \CLOPSh{} applies the same philosophy to
utility-scale workloads that \CLOPSv{} applied to QV-scale workloads.

This alignment enables a clearer performance narrative: layer fidelity
quantifies how well a system can execute a layer; \CLOPSh{} quantifies how fast
it can do so. Together, these benchmarks provide complementary perspectives on
the feasibility of large-scale, error-mitigated computation without conflating
speed and quality.

\textbf{Note on numerical comparability.} The adoption of utility-scale
hardware-aware circuits produces \CLOPSh{} values that differ from \CLOPSv{}
values measured on the same hardware. These differences reflect both genuine
speed improvements and differences in circuit definition (width, layer
structure, compilation model). \CLOPSv{} and \CLOPSh{} values should not be
compared on a common scale or plotted on the same axis without explicit
normalization and context.

\section{\label{sec:definition}Formal Definition of \texorpdfstring{\CLOPSh{}}{CLOPS\_h}}

\subsection{Terminology and Layer Abstraction}

We use the following terms precisely throughout this section.

A \textit{physical layer} is a set of qubit-disjoint two-qubit gates that can
be fully executed in parallel by the hardware during a single physical execution
interval. It is the atomic counting unit for \CLOPSh{} throughput.

An \textit{\EPLG{} layer} (used in the layer-fidelity context) is a quality
specification object: it defines which interactions should be assessed together
for fidelity estimation. An \EPLG{} layer has no direct execution semantics and
is not counted in \CLOPSh{}. The physical sublayers that realize an \EPLG{}
layer in practice are identical in structure to \CLOPSh{} physical layers,
enabling coherent joint interpretation of speed and quality measurements
obtained under the same operating conditions.


\CLOPSh{} and \LF{} intentionally share a layer decomposition but measure
different quantities: \LF{} measures how faithfully a layer executes; \CLOPSh{}
measures the rate at which layers execute.


\subsection{Physical Layer Definition}

Let $\mathcal{Q} = \{q_1, \ldots, q_N\}$ denote a connected set of $N$
physical qubits on a quantum processor, with native two-qubit connectivity
specified by a hardware coupling graph $G = (\mathcal{Q}, E)$.

A \CLOPSh{} physical layer $L$ consists of:

\begin{enumerate}
\item A set of two-qubit gates $\{U_{ij}\}$ for $(i,j)$ in a qubit-disjoint
  matching $M \subseteq E$, such that no qubit participates in more than one
  two-qubit gate in $L$.

\item Fully twirled single-qubit gates $\{V_i\}$ applied to all qubits in
  $\mathcal{Q}$, including those idle with respect to the two-qubit gates.
  These gates are treated as fixed per-layer overhead and do not alter the
  counting of physical layers.

\item A global synchronization barrier enforcing that all operations in $L$
  complete before execution of the subsequent layer begins.
\end{enumerate}

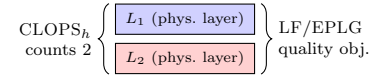
\begin{figure*}[t]
\centering
\tikzset{
    qubit/.style={circle, draw=black, fill=gray!15, minimum size=4.6mm, inner sep=0pt, font=\scriptsize},
    onqubit/.style={circle, draw=black, fill=blue!18, minimum size=4.6mm, inner sep=0pt, font=\scriptsize},
    idleq/.style={circle, draw=black!45, fill=white, minimum size=4.6mm, inner sep=0pt, font=\scriptsize, text=black!55},
    l1gate/.style={very thick, blue!70!black},
    l2gate/.style={very thick, red!70!black},
    faded/.style={draw=black!20, thin},
    caplbl/.style={font=\footnotesize},
}
\newcommand{\panelht}{3.6cm}

\begin{tabular}{@{}p{0.31\textwidth}@{\hspace{0.02\textwidth}}p{0.32\textwidth}@{\hspace{0.02\textwidth}}p{0.31\textwidth}@{}}

\begin{minipage}[t][\panelht][t]{0.31\textwidth}\centering
{\textbf{(a)}\ \footnotesize Heavy-hex device: select a connected 1D chain\par}
\vfill
\begin{tikzpicture}[scale=0.8, every node/.style={transform shape}]
  \foreach \x in {0,1,2,3,4} {
    \node[qubit, faded] (T\x) at (\x*0.95, 1.25) {};
    \node[qubit, faded] (B\x) at (\x*0.95, 0.0) {};
  }
  \foreach \x [evaluate=\x as \nextx using int(\x+1)] in {0,1,2,3} {
    \draw[faded] (T\x) -- (T\nextx);
    \draw[faded] (B\x) -- (B\nextx);
  }
  \foreach \x in {0,2,4} { \draw[faded] (T\x) -- (B\x); }
  \node[onqubit] (c0) at (0*0.95,1.25) {$q_0$};
  \node[onqubit] (c1) at (1*0.95,1.25) {$q_1$};
  \node[onqubit] (c2) at (2*0.95,1.25) {$q_2$};
  \node[onqubit] (c3) at (2*0.95,0.0)  {$q_3$};
  \node[onqubit] (c4) at (3*0.95,0.0)  {$q_4$};
  \node[onqubit] (c5) at (4*0.95,0.0)  {$q_5$};
  \draw[very thick, black] (c0)--(c1)--(c2)--(c3)--(c4)--(c5);
\end{tikzpicture}
\vfill
{\footnotesize Selected chain (blue), $N$ qubits, drawn on the device coupling graph (faded).\par}
\end{minipage}
&

\begin{minipage}[t][\panelht][t]{0.32\textwidth}\centering
{\textbf{(b)}\ \footnotesize Chain $\to$ two disjoint physical layers\par}
\vfill
\begin{tikzpicture}[scale=0.8, every node/.style={transform shape}]
  \node[caplbl, anchor=east] at (-0.2,1.15) {$L_1$};
  \foreach \i in {0,...,5} { \node[qubit] (b1\i) at (\i*0.62,1.15) {}; }
  \draw[l1gate] (b10)--(b11);
  \draw[l1gate] (b12)--(b13);
  \draw[l1gate] (b14)--(b15);
  \node[caplbl, anchor=east] at (-0.2,0.0) {$L_2$};
  \foreach \i in {0,...,5} { \node[qubit] (b2\i) at (\i*0.62,0.0) {}; }
  \draw[l2gate] (b21)--(b22);
  \draw[l2gate] (b23)--(b24);
  \node[idleq] at (0*0.62,0.0) {};
  \node[idleq] at (5*0.62,0.0) {};
\end{tikzpicture}
\vfill
{\footnotesize Each row is one hardware time slice, i.e.\ one \CLOPSh{} counting unit.\par}
\end{minipage}
&

\begin{minipage}[t][\panelht][t]{0.31\textwidth}\centering
{\textbf{(c)}\ \footnotesize One \EPLG{} layer $=\{L_1; L_2\}$\par}
\vfill
\begin{tikzpicture}[scale=0.8, every node/.style={transform shape}]
  \node[draw, fill=blue!18, minimum width=2.3cm, minimum height=0.5cm, font=\scriptsize] (barL1) at (0,0.65) {$L_1$ (phys.\ layer)};
  \node[draw, fill=red!18,  minimum width=2.3cm, minimum height=0.5cm, font=\scriptsize] (barL2) at (0,0.0)  {$L_2$ (phys.\ layer)};
  \draw[decorate, decoration={brace, amplitude=4pt, mirror}]
    ($(barL1.north west)+(-0.1,0)$) -- ($(barL2.south west)+(-0.1,0)$)
    node[midway, left=5pt, caplbl, align=right] {\CLOPSh{}\\counts $2$};
  \draw[decorate, decoration={brace, amplitude=4pt}]
    ($(barL1.north east)+(0.1,0)$) -- ($(barL2.south east)+(0.1,0)$)
    node[midway, right=5pt, caplbl, align=left] {\LF{}/\EPLG{}\\quality obj.};
\end{tikzpicture}
\vfill
{\footnotesize Same units, two roles: speed count vs.\ quality object.\par}
\end{minipage}

\end{tabular}
\caption{\label{fig:decomposition}Layer decomposition on a heavy-hex device and
its dual role. \textbf{(a)}~Both \LF{} and \CLOPSh{} are measured on a connected
$1$D chain of $N$ qubits (blue) selected from the device coupling graph (faded).
\textbf{(b)}~Because each physical layer must be a qubit-disjoint matching of
two-qubit gates, the chain decomposes into two disjoint physical layers: $L_1$
on the even bonds (starting at $q_0$) and $L_2$ on the odd bonds (starting at
$q_1$), following McKay~et~al.~\cite{mckay2023}. Qubits idle in a given slice
(open circles) still receive twirled single-qubit gates. Each physical layer is
one hardware time slice and one \CLOPSh{} counting unit. \textbf{(c)}~The ordered
pair $\{L_1; L_2\}$ constitutes one \EPLG{} layer: \LF{} assesses the pair as a
single quality object, while \CLOPSh{} counts the two physical layers as its
throughput unit---identical structure, different role. A $1$D chain requires
exactly two disjoint layers; general topologies require no more than the graph
degree plus one (Vizing's theorem), so the sublayer count is topology-dependent
while the physical layer itself remains a universal unit.}
\end{figure*}

The two-qubit gate set in each physical layer is one sublayer of the
\LF{}/\EPLG{} decomposition of $\mathcal{Q}$; a single physical layer does not
span the full coupling graph for any non-trivial device topology. Full
coupling-graph coverage requires multiple consecutive physical layers. When no
companion \LF{}/\EPLG{} decomposition is available, the maximally simultaneous
qubit-disjoint sets are used. Figure~\ref{fig:decomposition} illustrates this
decomposition and its dual role as the \CLOPSh{} counting unit and the \LF{}
quality object.

\subsection{Circuit Template}

A \CLOPSh{} circuit template is a parameterized quantum circuit across $N$ qubits 
and of depth $D$
physical layers:
\begin{equation}
  C(\boldsymbol{\theta}) = L_1(\boldsymbol{\theta}_1)\, L_2(\boldsymbol{\theta}_2) \cdots L_D(\boldsymbol{\theta}_D),
\end{equation}

where  $L_k(\boldsymbol{\theta}_k)$ is 
parameterized by $\boldsymbol{\theta}_k$. The \emph{structure} of layer $k$ is drawn from the
fixed sublayer decomposition of the qubit set (Fig.~\ref{fig:decomposition}b)
and repeats cyclically. For the $1$D chain, whose decomposition is the two
sublayers $L_1$ (even bonds) and $L_2$ (odd bonds), the depth sequence is
$L_1, L_2, L_1, L_2, \ldots$; a general $n$-sublayer topology cycles
$L_1 \ldots L_n$ likewise. The topology and structure of all physical layers are
fixed; only the single-qubit gate parameter values $\boldsymbol{\theta}_k$ vary
between depth positions and between executions.

The canonical benchmark operates on a template of width $N = 100$ and depth
$D = 100$ physical layers; these values, together with the shot count and the
convention for reporting other operating points, are specified in the Reporting
Requirements (Sec.~\ref{sec:definition}).

\begin{figure*}[t]
\centering
\tikzset{
    wire/.style={thick, black},
    vgate/.style={draw=black, fill=orange!20, rounded corners=1pt, minimum width=6.5mm, minimum height=5mm, inner sep=1pt, font=\scriptsize},
    ugate/.style={draw=black, fill=blue!16, rounded corners=1pt, minimum width=6.5mm, inner sep=1pt, font=\scriptsize},
    barrier/.style={dashed, thick, black!60},
    meas/.style={draw=black, fill=gray!12, minimum width=6.5mm, minimum height=5mm, inner sep=1pt, font=\scriptsize},
    reset/.style={draw=black, fill=gray!12, rounded corners=1pt, minimum width=6.5mm, minimum height=5mm, inner sep=1pt, font=\scriptsize},
    lbox/.style={draw=black, fill=blue!10, rounded corners=1.5pt, minimum width=8mm, minimum height=8mm, font=\scriptsize},
    lboxA/.style={lbox, fill=blue!16},
    lboxB/.style={lbox, fill=red!16},
    caplbl/.style={font=\footnotesize},
}

\begin{tabular}{@{}p{0.40\textwidth}@{\hspace{0.03\textwidth}}p{0.53\textwidth}@{}}

\begin{minipage}[t]{0.40\textwidth}\centering
{\textbf{(a)}\ \footnotesize Anatomy of one physical layer $L_1(\boldsymbol{\theta}_1)$\par}
\vspace{5mm}\par
{%
\begin{tikzpicture}[scale=1.15, every node/.style={transform shape}]
  \useasboundingbox (-0.55,0.7) rectangle (4.85,-3.6);
  \foreach \i in {0,...,5} {
    \node[caplbl, anchor=east] at (-0.15, -\i*0.62) {$q_\i$};
    \draw[wire] (0,-\i*0.62) -- (4.7,-\i*0.62);
  }
  \foreach \i in {0,...,5} {
    \node[vgate] at (0.75,-\i*0.62) {$V_{\i}$};
  }
  \foreach \a/\b in {0/1,2/3,4/5} {
    \draw[wire] (2.35,-\a*0.62) -- (2.35,-\b*0.62);
    \node[ugate, minimum height=1.2cm] at (2.35,{-(\a+\b)/2*0.62}) {$U_{\a\b}$};
  }
  \draw[barrier] (3.55,0.42) -- (3.55,-5*0.62-0.42);
  \node[caplbl, anchor=south, text=black!60] at (3.55,0.46) {barrier};
  \draw[-{Latex[length=2mm]}, black!55] (4.05,-2.5*0.62) -- (4.75,-2.5*0.62);
  \node[caplbl, anchor=west, text=black!55] at (4.55,-2.5*0.62-0.3) {$L_2$};
\end{tikzpicture}%
}
\end{minipage}
&

\begin{minipage}[t]{0.53\textwidth}\centering
{\textbf{(b)}\ \footnotesize Depth-$D$ template $C(\boldsymbol{\theta})$: structural layers cycled\par}
\vspace{5mm}\par
{%
\begin{tikzpicture}[scale=1.2, every node/.style={transform shape}]
  \useasboundingbox (-0.7,1.45) rectangle (7.1,-3.55);
  \foreach \i in {0,1,2} {
    \draw[wire] (0,-\i*0.9) -- (7.0,-\i*0.9);
  }
  \node[caplbl, anchor=east] at (-0.1,0) {$q_0$};
  \node[caplbl, anchor=east] at (-0.1,-0.9) {$\vdots$};
  \node[caplbl, anchor=east] at (-0.1,-1.8) {$q_{N\text{-}1}$};
  \foreach \i in {0,1,2} { \node[reset] at (0.55,-\i*0.9) {$|0\rangle$}; }
  \node[caplbl, anchor=south] at (0.55,0.55) {reset};
  \foreach \x/\sty/\lab/\th in {%
      1.5/lboxA/{$L_1$}/{$\boldsymbol{\theta}_1$},
      2.5/lboxB/{$L_2$}/{$\boldsymbol{\theta}_2$}} {
    \foreach \i in {0,1,2} { \node[\sty] at (\x,-\i*0.9) {\lab}; }
    \draw[barrier] (\x+0.5,0.5) -- (\x+0.5,-1.8-0.5);
    \node[caplbl, anchor=north, font=\scriptsize] at (\x,-1.8-0.55) {\th};
  }
  \node[caplbl] at (3.5,-0.9) {$\cdots$};
  \node[caplbl, anchor=north, font=\scriptsize] at (3.5,-1.8-0.55) {$\cdots$};
  \foreach \x/\sty/\lab/\th in {%
      4.5/lboxA/{$L_1$}/{$\boldsymbol{\theta}_{D\text{-}1}$},
      5.5/lboxB/{$L_2$}/{$\boldsymbol{\theta}_{D}$}} {
    \foreach \i in {0,1,2} { \node[\sty] at (\x,-\i*0.9) {\lab}; }
    \draw[barrier] (\x+0.5,0.5) -- (\x+0.5,-1.8-0.5);
    \node[caplbl, anchor=north, font=\scriptsize] at (\x,-1.8-0.55) {\th};
  }
  \foreach \i in {0,1,2} { \node[meas] at (6.5,-\i*0.9) {\scriptsize$\measuredangle$}; }
  \node[caplbl, anchor=south] at (6.5,0.55) {meas.};
  \draw[decorate, decoration={brace, amplitude=3pt}]
    (1.1,0.62) -- (2.9,0.62)
    node[midway, above=2pt, caplbl, font=\scriptsize, align=center] {repeat unit\\($=1$ \EPLG{} layer)};
  \draw[decorate, decoration={brace, amplitude=4pt, mirror}]
    (1.1,-1.8-1.05) -- (6.0,-1.8-1.05)
    node[midway, below=2pt, caplbl, font=\scriptsize] {$D$ physical layers ($k=1\ldots D$)};
\end{tikzpicture}%
}
\end{minipage}

\end{tabular}
\bigskip
\caption{\label{fig:anatomy}Executable structure of a \CLOPSh{} circuit.
\textbf{(a)}~One physical layer $L_k(\boldsymbol{\theta}_k)$ comprises fully
twirled single-qubit gates $V_i$ on all $N$ qubits (idle qubits included),
qubit-disjoint two-qubit gates $U_{ij}$ over a matching of the coupling graph,
and a terminating global synchronization barrier. \textbf{(b)}~The template
$C(\boldsymbol{\theta}) = L_1 L_2 \cdots L_D$ is built by \emph{cycling}
the fixed sublayer decomposition of Fig.~\ref{fig:decomposition}b---for the
chain, the pair $\{L_1, L_2\}$ (one \EPLG{} layer, the repeat unit)---to a total
of $D$ physical layers. Structural labels $L_1, L_2$ therefore recur along the
depth axis, but each depth position $k = 1 \ldots D$ binds its own parameter set
$\boldsymbol{\theta}_k$, so no two slices are identical even where their
structure matches. Each layer is closed by a barrier that forbids gate
reordering across layer boundaries, and the circuit is bracketed by reset and
measurement, both inside the \CLOPSh{} timing window. Whereas
Fig.~\ref{fig:decomposition} establishes the layer's dual role as speed unit and
quality object, this figure shows what the hardware actually executes within and
across layers.}
\end{figure*}
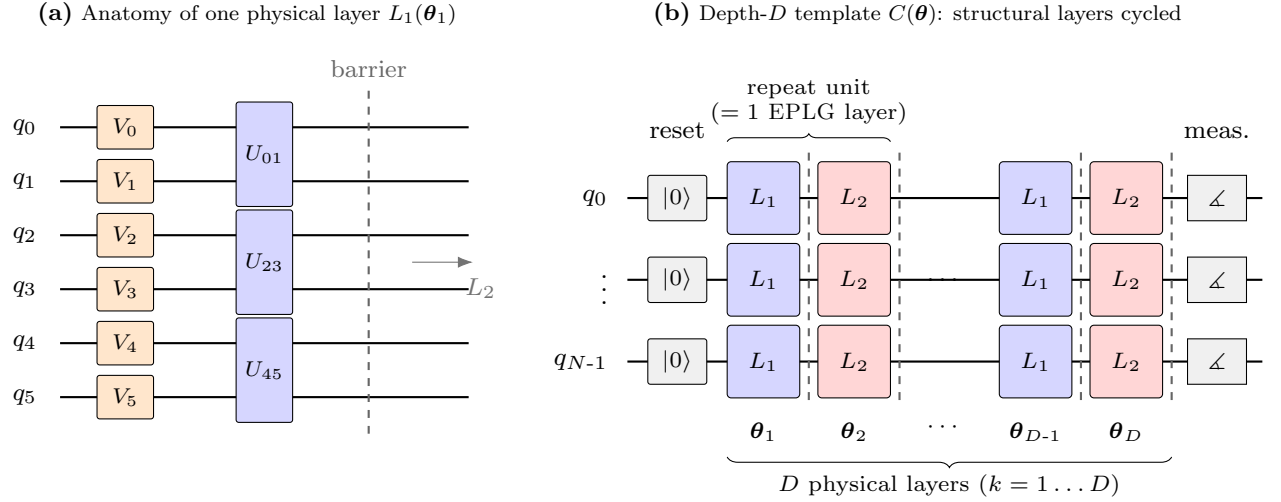

Typical parameterizations include single-qubit Clifford or Pauli twirling, and
hardware-native parameterized single-qubit rotations. The circuit template is
fixed at the beginning of the benchmark run and reused throughout.
The internal structure of a single physical layer and its composition into a
depth-$D$ template are shown in Fig.~\ref{fig:anatomy}.

\subsection{Parameterized Execution Model}

\CLOPSh{} models workloads in which the same circuit structure is executed
repeatedly with different parameter values. Let
$\{\boldsymbol{\theta}^{(1)}, \boldsymbol{\theta}^{(2)}, \ldots,
\boldsymbol{\theta}^{(K)}\}$ denote a sequence of $K$ independently generated
parameter sets applied to the fixed circuit template $C$, indexed by
$j = 1 \ldots K$. Each parameter set $\boldsymbol{\theta}^{(j)}$ is one full
circuit's worth of parameters---that is, the depth-indexed vector
$\boldsymbol{\theta}^{(j)} = (\boldsymbol{\theta}^{(j)}_1, \ldots,
\boldsymbol{\theta}^{(j)}_D)$ whose block $\boldsymbol{\theta}^{(j)}_k$
parameterizes physical layer $k$ of execution $j$. 

Key properties:

\begin{itemize}
\item Parameter updates do not depend on prior measurement outcomes.
\item The parameter stream is treated as effectively unbounded.
\item Each parameter instantiation corresponds to a distinct circuit execution
  including reset at circuit start and measurement at circuit end; both are
  included in the timing.
\item Each execution collects $S$ shots. The shot count $S$ is fixed across the
  parameter stream, so it enters the throughput measurement as a constant
  multiplier per execution.
\end{itemize}

This model captures execution patterns found in variational algorithms, layered
mitigation schemes, and large-scale calibration workloads. The canonical shot
count $S = 100$ is chosen as a representative midpoint between low-shot workloads
(e.g.\ expectation-value estimation with aggressive error mitigation) and
high-shot workloads (e.g.\ sampling-based methods), balancing the per-execution
fixed costs against per-shot measurement time; as with $N$ and $D$, other values
may be reported explicitly (Sec.~\ref{sec:definition}).

\subsection{Benchmark Execution Protocol}

\begin{enumerate}
\item \textit{Qubit set selection.} Select the $N$-qubit set $\mathcal{Q}$ to
  be benchmarked. When a layer-fidelity or \EPLG{} measurement is available,
  $\mathcal{Q}$ must be the same qubit set used for that measurement (e.g., the
  best 100-qubit chain identified during \EPLG{} characterization). When no
  such measurement is available, $\mathcal{Q}$ may be selected automatically
  using a topology-only heuristic (e.g., longest connected path in the coupling
  graph); this must be declared in reporting. \CLOPSh{} intentionally delegates
  quality-based qubit selection to the \LF{} workflow rather than duplicating
  quality optimization internally. A consequence of this delegation is that,
  when a companion \LF{}/\EPLG{} measurement exists, the benchmark width $N$ is
  not an independent \CLOPSh{} parameter but is fixed to the qubit-set width of
  that measurement. This is an internal-consistency requirement: it prevents a
  speed number measured at one width from being paired with a quality claim
  established at another.

\item \textit{Layer decomposition.} Compute the canonical physical layer
  decomposition for $\mathcal{Q}$. This decomposition must match the one used
  in the corresponding \LF{}/\EPLG{} measurement.

\item \textit{Circuit preparation.} Construct the parameterized circuit template
  $C(\boldsymbol{\theta})$ of width $N$ and depth $D$ physical layers, and
  compile it to a parameterized representation compatible with the target
  backend.

\item \textit{Parameter stream generation.} Generate a sequence of parameter
  sets $\{\boldsymbol{\theta}^{(j)}\}_{j=1}^{K}$ using a pseudo-random process.

\item \textit{Pipeline saturation.} Submit parameterized circuit executions
  continuously with S shots per circuit. The first result batch returned by the system marks the end of
  the pipeline-fill transient. The execution that produced this first result is
  excluded from all throughput calculations. In practice, a clear separation
  between the transient and steady-state regimes is observed: execution rate
  stabilizes rapidly once the pipeline is filled, making the first-result-batch
  boundary a reliable and observable transition point.

\item \textit{Steady-state measurement.} The measurement interval begins at the
  timestamp of the first result batch and ends at the timestamp of the final
  result batch. Let $t_1$ be the wall-clock time at which the first result
  batch is received, and $t_n$ be the time at which the $n$-th (final) result
  batch is received. Then:
  \begin{equation}
    T_{\mathrm{ss}} = t_n - t_1.
  \end{equation}
  $K_{\mathrm{ss}}$ counts the circuit executions completed in result batches 2
  through $n$ (inclusive). \CLOPSh{} is then:
  \begin{equation}
    \CLOPSh{} = \frac{K_{\mathrm{ss}} \cdot D \cdot S}{T_{\mathrm{ss}}}.
    \label{eq:clops_h}
  \end{equation}
  \CLOPSh{} is reported in physical layers per second. It characterizes hardware
  parallelism, control-stack throughput, and barrier-delimited scheduling
  performance. Because each physical layer corresponds to one time slice of an
  $N$-qubit parallel operation, \CLOPSh{} directly bounds the execution rate of
  layered $N$-qubit circuits: a device sustaining $R$ physical layers per second
  can execute approximately $R/(D \cdot S)$ \CLOPSh{} type circuits per second. 

  This definition is deliberately implementation-agnostic: it does not require
  the implementer to detect convergence or characterize pipeline depth. The
  first-result boundary is observable at the user level on any execution
  architecture.

\item \textit{Accounting of execution cost.} The measured time must include all
  operations required to execute each parameterized instance, including
  parameter binding, runtime or post-bind compilation, payload construction and
  data movement, control electronics initialization and loading, QPU execution
  and measurement, and result retrieval. The protocol does not prescribe how
  executions are batched, pipelined, or scheduled, provided that all required
  costs are included.

  This rule fixes the boundary between \CLOPSh{} and software-stack
  benchmarking (Section~\ref{sec:evolution}): \CLOPSh{} counts exactly the
  circuit-preparation cost that is \emph{actually incurred per execution}. It is
  therefore not a fixed carve-out but a consequence of holistic, steady-state
  accounting. The one-time cost of constructing and transpiling the circuit
  template---the class of work measured by suites such as
  Benchpress~\cite{nation2025}---is not part of the steady-state
  measurement when it is amortized outside the timing window. On platforms using
  parametric compilation, where the template is compiled once and each parameter
  set requires only a lightweight substitution, this construction cost falls
  outside $T_{\mathrm{ss}}$ by construction and does not contribute to
  \CLOPSh{}. On platforms where every parameter set instead forces a full
  recompilation or re-routing pass, that same work is incurred per execution
  and is therefore naturally included in $T_{\mathrm{ss}}$. In both cases
  \CLOPSh{} reports the throughput a user actually experiences, without
  privileging any particular compilation strategy.
\end{enumerate}

\subsection{\label{sec:lfoc}Operating Conditions}

\CLOPSh{} is a speed benchmark, not a quality benchmark. However, because gate
duration, calibration state, and error suppression settings all affect both
speed and quality, an unconstrained speed benchmark can be trivially inflated
by degrading quality. To prevent this, \CLOPSh{} is defined to be executed
under \textit{layer-fidelity operating conditions} (\LFOC{}).

A \CLOPSh{} execution satisfies \LFOC{} if:

\begin{itemize}
\item \textit{Gate definitions and durations.} All single-qubit and two-qubit
  gates use the same native gate definitions, calibrated pulse implementations,
  and gate durations as those used in the corresponding \LF{} measurement. No
  gate parameter may be altered for \CLOPSh{} unless the same alteration is
  applied to the \LF{} measurement.

\item \textit{Scheduling constraints.} Barriers are enforced at all layer
  boundaries. Gates may be reordered within a physical layer but not across
  layer boundaries. Compilation policies are compatible with those used for
  \LF{} execution.

\item \textit{Error suppression policies.} Any error suppression mechanisms
  (e.g., dynamic decoupling, crosstalk mitigation) enabled during \LF{}
  execution must be equivalently enabled during \CLOPSh{} execution. \CLOPSh{}
  must not disable or weaken mechanisms that are active under \LF{} conditions.

\item \textit{Calibration consistency.} \CLOPSh{} and \LF{} must be executed
  within the same calibration cycle, or under calibrations explicitly declared
  equivalent. Maintaining separate fast-calibration and high-quality-calibration
  regimes for \CLOPSh{} and \LF{} respectively is not permitted.
\end{itemize}

\CLOPSh{} does not verify these conditions internally. The quality anchor is
external and must be declared as part of result reporting.

\subsection{Reporting Requirements}

A reported \CLOPSh{} result must specify:
\begin{itemize}
\item The circuit shape: number of qubits $N$ and circuit depth $D$
  (number of physical layers)
\item The execution parameters: shots $S$ per circuit execution, and the
  parameterization and twirling configuration
\item \CLOPSh{} value in physical layers per second
\item Whether the qubit set was externally specified (e.g., from an \EPLG{}
  measurement) or automatically selected; if the former, cite the corresponding
  \LF{}/\EPLG{} result
\item Declaration of layer-fidelity operating conditions compliance
\item Execution interface used (e.g., runtime primitives)
\end{itemize}

To support cross-system comparability, these free parameters require a shared
default. An unqualified ``\CLOPS{}'' number therefore denotes the \CLOPSh{}
result measured at the canonical operating point---circuit shape $N = D = 100$
and execution shot count $S = 100$---under \LFOC{}. Any result reported at a
different operating point must carry these parameters in the symbol itself,
written $\CLOPSh(N, D, S)$, so that the operating point travels with the number
rather than residing in a footnote.

\section{Case Study: \texorpdfstring{\CLOPSh{}}{CLOPS\_h} in Practice}

\subsection{Overview}

This section presents a case study illustrating how \CLOPSh{} behaves as a
progress-tracking and bottleneck-diagnostic tool across multiple years of system
development. The data are drawn from measurements
spanning approximately 3.5 years of hardware, firmware, and software
development on IBM Quantum systems, and are intended to demonstrate how
\CLOPSh{} can be used to track progress, decompose performance gains, and
identify shifting bottlenecks across the performance hierarchy.

The primary data are paired measurements of \CLOPSh{} and application-level
wall-clock runtime, obtained simultaneously on the same device under identical
operating conditions. The application benchmark is the Trotterized dynamics
experiment from Kim et al.~\cite{kim2023}---a well-characterized, utility-scale
circuit executed with fixed configuration as a strict speed benchmark which we 
refer to as TDE circuits (Trotterized Dynamics Experiment circuits). Circuit
structure, shot count, and error mitigation configuration were held constant
across all measurement points; only the underlying system was updated between
measurements. This design isolates speed improvements from quality changes and
treats the Kim et al.\ experiment as a fixed application-level probe, analogous
in intent to the SPEC benchmark philosophy at the application level of the
performance hierarchy. 

\subsection{\texorpdfstring{\CLOPSh{}}{CLOPS\_h} Improvements and Application-Level Gains}

Over the measurement period, \CLOPSh{} improved by approximately $330\times$
relative to the baseline. Paired application-level measurements show wall-clock
runtime improvements of approximately $75\times$ and $110\times$ at two
representative points in the trajectory, corresponding to \CLOPSh{} gains of
$250\times$ and $330\times$ respectively. Table~\ref{tab:results} summarizes
these measurements.

\begin{table}[h]
\caption{\label{tab:results}Summary of \CLOPSh{} and application-level
performance gains relative to the June 2022 baseline.}
\begin{ruledtabular}
\begin{tabular}{lccc}
\textbf{Date} & \textbf{\CLOPSh{}} & \textbf{\CLOPSh{} gain} & \textbf{App.\ runtime gain} \\
\hline
June 2022 & 950       & $1\times$ (baseline) & $1\times$ (baseline) \\
Mar.\ 2025 & 250{,}000 & $\sim\!250\times$    & $\sim\!75\times$     \\
Nov.\ 2025 & 330{,}000 & $\sim\!330\times$    & $\sim\!110\times$    \\
\end{tabular}
\end{ruledtabular}
\end{table}

In both cases, application-level gains are substantial but smaller than
\CLOPSh{} gains, reflecting costs at the application layer of the performance
hierarchy that \CLOPSh{} does not measure. This is expected by design: \CLOPSh{}
captures system-level layer throughput, while application runtime also includes
classical optimization orchestration, workflow management, and other overhead
that sits above the system level.

\subsection{Hardware and Software Contributions to \texorpdfstring{\CLOPSh{}}{CLOPS\_h} Gains}

The $250\times$ \CLOPSh{} improvement from June 2022 to March 2025 reflects
contributions from two distinct sources: hardware and device-level improvements,
and software stack optimization---principally parametric compilation.
Understanding their relative contributions requires combining device-level
metrics carefully, because neither \CPS{} nor gate time alone provides an
accurate picture.

For June 2022, the delay between circuits (rep delay) was approximately $1{,}000\,\mu\mathrm{s}$ and
state preparation plus measurement added roughly $12\,\mu\mathrm{s}$, giving a
\CPS{} ceiling of approximately 988. Gate time for the TDE circuit was
approximately $73\,\mu\mathrm{s}$, bringing total per-circuit time to roughly
$1{,}085\,\mu\mathrm{s}$ and an effective device-level circuit rate of
approximately 921 TDE circuits per second. For March 2025, rep delay had
dropped to $250\,\mu\mathrm{s}$ with state preparation and measurement
unchanged at $12\,\mu\mathrm{s}$, and gate time had fallen sharply to
approximately $12\,\mu\mathrm{s}$, giving a \CPS{} ceiling of approximately
3{,}816 and an effective device-level circuit rate of approximately 3{,}649
TDE circuits per second.

The ratio of these effective device-level rates---$3{,}649 / 921 \approx
4\times$---represents the hardware-driven improvement in the achievable ceiling
over this period. This accounts for roughly $4\times$ of the $250\times$ total
\CLOPSh{} gain; the remaining ${\sim}60\times$ reflects software and control
stack improvements, of which parametric compilation was the dominant
contributor.

This decomposition also illustrates a subtlety worth noting for practitioners
conducting their own performance characterization. \CPS{} alone ($988 \to
3{,}816$, ${\sim}3.9\times$) and gate time reduction alone ($73\,\mu\mathrm{s}
\to 12\,\mu\mathrm{s}$, ${\sim}6\times$) each suggest different magnitudes of
hardware improvement. Neither is wrong in isolation, but neither correctly
characterizes the ceiling relevant to a real workload. The effective
device-level circuit rate---combining rep delay, state preparation and
measurement, and gate time into a single per-circuit cost---is the quantity that
actually bounds system and application throughput.

For the software contribution, parametric compilation was one of the dominant
drivers of performance. In the June 2022 baseline configuration, each new
parameter set required a complete recompilation pass before QPU execution could
begin, leaving the QPU idle for the majority of wall-clock time. The transition
to end-to-end parametric compilation---in which the circuit structure is
compiled once and subsequent parameter updates require only a lightweight
substitution step---eliminated this bottleneck and enabled near-continuous QPU
utilization. Significant additional contributions came from improvements in
control electronics throughput, scheduling efficiency, and data movement latency
across the classical-quantum execution pipeline. \CLOPSh{} captures the
aggregate effect of all of these changes directly, by construction: systems
where classical gaps between QPU execution bursts remain large show lower
\CLOPSh{}, while systems with efficient parametric pipelines approach the limit
set by native gate and readout durations.

\subsection{Amdahl's Law and the Shifting Bottleneck}

From March 2025 to November 2025, device-level characteristics were essentially
unchanged: \CPS{}, gate time, and the effective device-level circuit rate all
remained constant at approximately their March 2025 values, giving a \CLOPSh{}
device ceiling of approximately 365{,}000 physical layers per second
($3{,}649\,\mathrm{circuits/s} \times 100\,\mathrm{layers/circuit}$). Against
this fixed ceiling, measured \CLOPSh{} improved from 250{,}000 to
330{,}000---from approximately 68\% to approximately 90\% of the device limit.
This second period of improvement was driven entirely by software and control
stack optimization, with no contribution from hardware changes.

The 68\% efficiency at March 2025 and the subsequent push to 90\% illustrates
why \CLOPSh{} is a useful diagnostic beyond raw performance reporting. At 68\%
of the device ceiling, substantial headroom remained in the software stack. By
November 2025, with the system approaching 90\% of what the hardware permits,
the device itself becomes the next natural optimization target---a conclusion
that would not be visible from device-level metrics alone.

An additional observation from these measurements is that the ${\sim}3:1$ ratio
between \CLOPSh{} gains and application-level gains did not arise from a fixed
structural overhead but from coordinated optimization across the full stack.
Following Amdahl's Law~\cite{amdahl1967}, continued improvement in the \CLOPSh{}-relevant portion
of the stack would have produced diminishing returns at the application level:
the non-\CLOPSh{} fraction, though small in absolute terms, would have come to
dominate application runtime. The team recognized this inflection and
deliberately pushed on application-layer overhead in parallel with continued
\CLOPSh{} improvements. The result was that the ${\sim}3:1$ ratio held in
November 2025 rather than worsening.

\subsection{Implications for the Performance Hierarchy}

These results demonstrate that \CLOPSh{} is a reliable leading indicator of
application-level speed improvement, but not a direct predictor of it.
System-level gains translate to application-level gains only to the extent that
application-layer overhead is also managed.

This coupling also illustrates why the application level of the performance
hierarchy (Table~\ref{tab:hierarchy}) requires its own benchmark. A principled
application-level speed benchmark faces an integrity problem that does not arise
at the system level. \CLOPSh{} is protected from speed-quality gaming by its
operating conditions requirement (Sec.~\ref{sec:lfoc}). No analogous anchor
currently exists at the application level. In the pre-advantage era, where classical computation is
legitimately part of quantum-classical workflows, a pure application speed
benchmark cannot easily distinguish a system that is genuinely accelerating
computation with quantum hardware from one that is routing to a classical solver.

This problem resolves naturally once quantum advantage is established for a
given problem class. At that point, an application-level speed benchmark can be
anchored to an advantage characteristic---whether superior speed, reduced cost,
or access to problem sizes that are classically intractable---in much the same
way that \CLOPSh{} is anchored to layer fidelity. Until then, the Kim et al.\
experiment used here represents a pragmatic interim approach: a concrete
demonstration of what application-level speed measurement could look like when
the quantum resource contribution is not in question. Formalizing this into a
hardware-agnostic, gaming-resistant application-level benchmark remains an
important open problem for the field.

Across all measurement points, \CLOPSh{} proves an effective instrument for
tracking system-level progress and diagnosing where bottlenecks reside in the
performance hierarchy. The $330\times$ improvement over two years
reflects contributions at every level of the stack---from hardware gate time and
rep delay reduction, to parametric compilation and control pipeline
optimization---and the paired application-level measurements confirm that
system-level gains translate to user-observable outcomes, modulo overhead at the
application layer.

\section{Discussion and Limitations}

\subsection{\texorpdfstring{\CLOPSh{}}{CLOPS\_h} as a Speed Metric, Not a Performance Guarantee}

\CLOPSh{} measures speed, not algorithmic success. A high \CLOPSh{} value
indicates that a system can execute layered circuits at a high sustained rate,
but does not imply that the results of those executions are correct, accurate,
or useful for a given application. Fidelity and error structure must be
characterized independently. \CLOPSh{} is best interpreted as a capacity metric
that quantifies how rapidly a system can carry out a given class of layered
operations tied to LFOC.

\subsection{Representativeness of the Circuit Family}

\CLOPSh{} circuits are designed to resemble the layered structures found in many
near-term and utility-scale algorithms, including variational circuits and error
mitigation workflows. Nonetheless, no single circuit family can represent the
full diversity of quantum workloads.

In particular, \CLOPSh{} does not capture:
\begin{itemize}
\item Algorithms with highly irregular or data-dependent control flow.
\item Circuits with extensive mid-circuit measurement and real-time classical
  feedback.
\item Applications whose gate topology requires complex routing or does not map
  naturally onto the canonical physical-layer decomposition.
\end{itemize}

\subsection{Dependence on Benchmark Parameters and Operating Conditions}

The numerical value of \CLOPSh{} depends on the specific choices of benchmark
parameters: number of qubits $N$, circuit depth $D$, shot count $S$,
parameterization scheme, and layer decomposition. While the benchmark definition
is unambiguous for any fixed set of parameters, comparisons across systems or
over time are meaningful only when these parameters are held constant or
explicitly documented.

The direction of these dependencies is predictable. As $N$ increases,
\CLOPSh{} may decrease, reflecting increased control orchestration, data
movement, and synchronization costs across a larger qubit set. As $D$
increases, \CLOPSh{} will generally improve, reflecting the amortization of
fixed per-circuit costs---qubit initialization, measurement, and inter-circuit
delays---across more layers; this effect plateaus once depth is sufficient to
render these fixed costs negligible relative to total layer execution time. The
default parameters of $N = 100$ qubits, $D = 100$ layers, and $S = 100$ shots were chosen to
reflect utility-scale workloads and to ensure these fixed costs are
representative of practical application conditions.

Because these dependencies are both strong and predictable in direction, they
create a comparability hazard for headline reporting: two systems could each
report a compliant \CLOPSh{} number yet differ by orders of magnitude solely
through circuit-shape selection---for example a small-$N$, large-$D$, high-$S$
configuration against the utility-scale default. This hazard is the rationale for
pinning an unqualified \CLOPS{} number to the canonical operating point
(Sec.~\ref{sec:definition}). The canonical point is not
expected to remain fixed indefinitely; as with the periodic re-baselining of
classical benchmark suites, successive utility-scale reference points are
expected to be defined as device capabilities advance, preserving comparability
within a reference generation while allowing the benchmark to track the regime
it is meant to characterize.

A related concern is the sensitivity of \CLOPSh{} to operating parameters that
affect both speed and quality, such as gate duration and calibration state. This
is precisely the hazard the layer-fidelity operating conditions requirement
(Sec.~\ref{sec:lfoc}) is designed to prevent. Cross-system comparisons are
therefore most meaningful when both \CLOPSh{} results and a corresponding
\LF{}/\EPLG{} measurement are reported together.

\subsection{Sensitivity to Software and Systems Design}

One of the strengths of \CLOPSh{} is its sensitivity to software and runtime
architecture. Because classical overheads are included by design, improvements
in compilation, parameter handling, scheduling, and control orchestration can
lead to substantial gains in \CLOPSh{} even when hardware gate performance
remains unchanged. \CLOPSh{} should therefore not be interpreted solely as a
measure of hardware capability; it reflects the combined performance of
hardware, firmware, runtime, and software stack. This holistic nature is
consistent with the benchmark's goals, but requires care when attributing
performance changes to specific system components.

\subsection{Using Metrics Across Hierarchy Levels}

The benchmark hierarchy of Table~\ref{tab:hierarchy} is not only a taxonomy 
but a diagnostic toolkit: metrics at different levels can be composed to 
characterize workloads that no single metric fully covers. Error-detecting 
circuits based on coherent Pauli checks (CPC)~\cite{vandenberg2023,gonzales2023,martiel2025} 
provide a concrete illustration.

CPC circuits execute a fixed layered circuit repeatedly, postselecting on ancilla 
flag qubits to reject shots in which errors are detected. For any given device, 
the postselection fraction is determined by circuit fidelity---a fixed quality 
characteristic of the hardware---meaning a substantial fraction of shots will be 
discarded regardless of how the experiment is run. Overcoming this overhead
requires executing a large number of shots in a practical time window.

\CPS{} and \CLOPSh{} address this throughput demand at complementary depth regimes.
For shallow CPC circuits, \CPS{}~[Eq.~\eqref{eq:mcps}] provides the relevant
device-level ceiling, bounding the raw shot rate that cannot be exceeded. As the circuits get deeper
---such as the spacetime code constructions of
Martiel and Javadi-Abhari~\cite{martiel2025}, which use hardware-aware circuits
and thousands of two-qubit gates---the expected performance moves from what \CPS{}
predicts to what would be indicated by \CLOPSh{}, 
capturing payload loading, pipeline fill, and sustained execution rate under 
realistic operating conditions. 

\subsection{\label{sec:future-scope}Future Scope and Applicability}

\begin{table*}[t]
\caption{\label{tab:framework-evolution}Evolution of the \CLOPS{} framework.
\CLOPS{} has consistently measured the sustained throughput of an interactive
quantum computing system under operating conditions known to produce meaningful
results, at the level of abstraction users and algorithms care about. At each
transition---from QV-scale to utility-scale, from utility-scale to early
fault-tolerant, and from early fault-tolerant to full fault-tolerant---the
counting unit, quality anchor, and characteristic bottlenecks shift together.
Each column is a fresh instantiation of the same design pattern, not a
refinement of the previous one. The shift in interpretation from circuit
throughput (columns 1, 2, and 4) to error-correction engine throughput (column
3) reflects what users care about in each era: in the early fault-tolerant
regime, the user-relevant question is how capable the QEC engine is, not yet how
fast logical algorithms run. The two rightmost columns indicate the expected
direction of framework evolution and do not assert completed specifications.}
\begin{ruledtabular}
\begin{tabular}{p{3.4cm}p{3.3cm}p{3.4cm}p{3.3cm}p{3.3cm}}
 & \textbf{Small Scale} & \textbf{Utility Scale} & \textbf{Early FT} & \textbf{Full FT} \\
\hline
\textbf{Quality anchor}      & QV                & Layer fidelity       & Logical error rate         & Logical gate fidelity \\[2pt]
\textbf{Execution structure} & QV circuits       & Layered circuits     & Syndrome cycles           & Logical circuits \\[2pt]
\textbf{Counting unit}       & QV layers         & Physical layers      & Syndrome cycles           & Logical layers \\[2pt]
\textbf{Key bottlenecks}     & Gate fidelity, compilation & Parametric compilation, classical pipeline & Syndrome cycle rate & Magic state factories, T-gate cost, distillation throughput, decoding latency\\[2pt]
\textbf{Interpretation}      & QV-scale circuit throughput & Utility-scale circuit throughput & Error-correction engine throughput & Logical circuit throughput \\
\end{tabular}
\end{ruledtabular}
\end{table*}

As quantum systems evolve from error-mitigated execution toward fault-tolerant
regimes, the \CLOPS{} framework is expected to evolve twice more
(Table~\ref{tab:framework-evolution})---each time following the same design
principle that distinguished \CLOPSv{} from \CLOPSh{}: adopt the counting unit
that users and algorithms care about at that level of the stack, and anchor the
speed measurement to the quality benchmark at the same level.

The first expected transition is to syndrome cycles. In early fault-tolerant systems,
the operation users and the hardware care about is the syndrome extraction cycle:
a fixed sequence of physical operations executed continuously to track 
errors. This is the counting unit users reason about when assessing
whether their error-correction protocol is keeping pace with the computation.
A \CLOPS{} variant in this regime would measure the sustained rate of syndrome
cycle execution, anchored to an independently verified syndrome or logical
fidelity benchmark. The characteristic bottlenecks shift accordingly---from the
classical pipeline and parametric compilation latency that dominate the utility
era, to syndrome cycle rate. \CLOPSh{} remains a
useful architectural characterization of the physical layer, but it is no longer
the benchmark users reach for to assess system speed.

The second transition is to logical layers. Once fault-tolerant systems mature,
users and algorithms operate at the abstraction of the logical layer: parallel
executions of encoded gates across the logical qubit register. The relevant
throughput question becomes how quickly the system can sustain parallel logical
gate execution, and the bottlenecks shift again. Magic state factory throughput,
distillation latency, and the ratio of T-gate cost to Clifford cost become the
diagnostic quantities---exactly analogous to how parametric compilation and
control pipeline latency were the diagnostic quantities in the utility era, and
decoding latency in the early fault-tolerant era. A \CLOPS{} variant here would
measure sustained logical layers per second, anchored to a logical gate fidelity
or code-distance quality benchmark.

The continuity across these variants is one of design philosophy, not metric
hierarchy. Each \CLOPS{} variant is a fresh instantiation of the same pattern:
measure the sustained throughput of the counting unit users and algorithms care
about at that level of the stack; anchor the measurement to the quality benchmark
at the same level; include all classical and quantum costs holistically; and
target steady-state operation so that the result reflects sustained workload
performance rather than transient peaks. What changes across eras is which
counting unit is user-relevant and which bottlenecks dominate---not the
underlying measurement discipline. The \CLOPS{} framework is designed to track
this progression gracefully: each variant serves as the primary user-facing speed
benchmark for its era, then yields to the next instantiation as the relevant
abstraction level shifts. What ``user-facing'' means shifts too: in the early
fault-tolerant era, the user-relevant question is not yet algorithmic
throughput but QEC infrastructure readiness---how capable is the error-correction
engine, and how close is the system to sustaining logical computation. The
framework returns to algorithmic throughput at the logical layer, completing the
arc from QV-scale circuits through utility-scale circuits to logical circuits,
with QEC infrastructure throughput as the necessary bridge between them.

\section{Conclusion}

As quantum computing systems enter the utility era, execution speed has become a
decisive factor in determining the feasibility of large-scale, error-mitigated
quantum experiments. In this work, we have formalized \CLOPSh{} as a holistic
speed benchmark designed explicitly for this regime. \CLOPSh{} measures the
sustained rate at which a system executes parallel two-qubit gate slices per
second---physical layers---directly connecting to published device capability
claims about circuit size and characterizing whether target utility-scale
workloads are feasible within practical time constraints.

A key contribution of \CLOPSh{} is its structural alignment with layer-fidelity
benchmarking. The physical layers used as \CLOPSh{}'s counting unit are the same
sublayers that realize an \EPLG{} layer in practice, so speed and quality
measurements share a common circuit structure and can be interpreted jointly
without conflating the two---\CLOPSh{} counting the layers as a throughput unit,
\LF{} assessing them as a quality object. Grounding \CLOPSh{} in this shared
decomposition also keeps its scope explicit, avoiding overclaimed applicability
to future fault-tolerant logical-layer metrics, where different counting units
will be required.

A key emphasis of \CLOPSh{} is its explicit operating conditions requirement. At
the qubit level, speed and quality are implicitly co-measured---gate times are
reported for calibrated, quality-characterized gates, and no one reports the
speed of a gate they have not also tuned up. This implicit coupling is so
natural at the qubit level that it is rarely stated. At the system level it
breaks down: speed and quality benchmarks are independent measurements that can
diverge, and the gap between them creates room for optimization that inflates
one at the expense of the other. The original \CLOPS{} benchmark recognized this
problem and addressed it by tying speed measurement to \QV{} operating
conditions---a principle that has proven difficult to enforce in practice as
devices scaled beyond the \QV{} regime and the implicit anchor weakened~\cite{wack2021}.
\CLOPSh{} restores and formalizes this coupling by requiring execution under
layer-fidelity operating conditions, explicitly declared and tied to an
independently measured quality benchmark. Speed and quality remain independently
reported, but are anchored to the same physical operating envelope---restoring
at the system level what has always been assumed at the qubit level.

Together, these properties make \CLOPSh{} a foundation for tracking progress,
diagnosing bottlenecks, and guiding system co-design across the performance
hierarchy. The results presented here---$330\times$ gains in \CLOPSh{} over two
years, translating to $110\times$ gains in application-level wall-clock
runtime---demonstrate both the practical value of the benchmark and the
importance of optimizing at every level of the hierarchy independently.

Looking forward, the \CLOPS{} framework points toward two open problems. The
first is an application-level speed benchmark: one that can do for the
application tier what \CLOPSh{} does for the system tier. As discussed in
Section~\ref{sec:evolution}, this benchmark faces a fundamental integrity
problem in the pre-advantage era---a pure application speed benchmark cannot
easily distinguish a system that is genuinely accelerating computation with
quantum hardware from one that is routing to a classical solver. This problem
resolves once quantum advantage is established for a given problem class, at
which point an application-level speed benchmark can be anchored to an advantage
characteristic in the same way \CLOPSh{} is anchored to layer fidelity.

The second open problem is the evolution of \CLOPS{} itself into the
fault-tolerant era. As detailed in Section~\ref{sec:future-scope}, the same design
pattern that defines \CLOPSh{} anticipates further instantiations for the
syndrome-cycle and logical-layer regimes---each a fresh re-instantiation of the
framework rather than a refinement, just as \CLOPSh{} is a re-instantiation of
\CLOPSv{} for a new regime. \CLOPSh{} is the current instantiation; its value
lies not only in what it measures today, but in the framework it establishes for
measuring what comes next.

\bibliography{references}

@misc{wack2021,
      title={Quality, Speed, and Scale: three key attributes to measure the performance of near-term quantum computers}, 
      author={Andrew Wack and Hanhee Paik and Ali Javadi-Abhari and Petar Jurcevic and Ismael Faro and Jay M. Gambetta and Blake R. Johnson},
      year={2021},
      eprint={2110.14108},
      archivePrefix={arXiv},
      primaryClass={quant-ph},
      url={https://arxiv.org/abs/2110.14108}, 
}

@article{kandala2017,
  author  = {Kandala, Abhinav and Mezzacapo, Antonio and Temme, Kristan and Takita, Maika and Brink, Markus and Chow, Jerry M. and Gambetta, Jay M.},
  title   = {Hardware-efficient Variational Quantum Eigensolver for Small Molecules and Quantum Magnets},
  journal = {Nature},
  volume  = {549},
  pages   = {242--246},
  year    = {2017},
  doi     = {10.1038/nature23879}
}

@misc{farhi2014,
      title={A Quantum Approximate Optimization Algorithm}, 
      author={Edward Farhi and Jeffrey Goldstone and Sam Gutmann},
      year={2014},
      eprint={1411.4028},
      archivePrefix={arXiv},
      primaryClass={quant-ph},
      url={https://arxiv.org/abs/1411.4028}, 
}

@article{temme2017,
  author  = {Temme, Kristan and Bravyi, Sergey and Gambetta, Jay M.},
  title   = {Error Mitigation for Short-Depth Quantum Circuits},
  journal = {Physical Review Letters},
  volume  = {119},
  pages   = {180509},
  year    = {2017},
  doi     = {10.1103/PhysRevLett.119.180509}
}

@article{cross2019,
  author  = {Cross, Andrew W. and Bishop, Lev S. and Sheldon, Sarah and Nation, Paul D. and Gambetta, Jay M.},
  title   = {Validating Quantum Computers Using Randomized Model Circuits},
  journal = {Physical Review A},
  volume  = {100},
  pages   = {032328},
  year    = {2019},
  doi     = {10.1103/PhysRevA.100.032328}
}

@article{endo2018,
  author  = {Endo, Suguru and Benjamin, Simon C. and Li, Ying},
  title   = {Practical Quantum Error Mitigation for Near-Future Applications},
  journal = {Physical Review X},
  volume  = {8},
  pages   = {031027},
  year    = {2018},
  doi     = {10.1103/PhysRevX.8.031027}
}

@inproceedings{murali2020,
  author    = {Murali, Prakash and Baker, Jonathan M. and Javadi-Abhari, Ali and Chong, Frederic T. and Martonosi, Margaret},
  title     = {Noise-Adaptive Compiler Mappings for Noisy Intermediate-Scale Quantum Computers},
  booktitle = {Proceedings of the 25th International Conference on Architectural Support for Programming Languages and Operating Systems (ASPLOS)},
  year      = {2019},
  doi       = {10.1145/3297858.3304075}
}

@misc{mckay2023,
      title={Benchmarking Quantum Processor Performance at Scale}, 
      author={David C. McKay and Ian Hincks and Emily J. Pritchett and Malcolm Carroll and Luke C. G. Govia and Seth T. Merkel},
      year={2023},
      eprint={2311.05933},
      archivePrefix={arXiv},
      primaryClass={quant-ph},
      url={https://arxiv.org/abs/2311.05933}, 
}

@article{dongarra2003,
  author  = {Dongarra, Jack J. and Luszczek, Piotr and Petitet, Antoine},
  title   = {The {LINPACK} Benchmark: Past, Present, and Future},
  journal = {Concurrency and Computation: Practice and Experience},
  volume  = {15},
  pages   = {803--820},
  year    = {2003},
  doi     = {10.1002/cpe.728}
}

@misc{merz2026,
      title={Crossing the 12,000-atom barrier with heterogeneous quantum-classical supercomputing: quantum chemistry of protein-ligand complexes}, 
      author={Merz, Jr., Kenneth M. and Akhil Shajan and Danil Kaliakin and Fangchun Liang and Yuichi Otsuka and Tomonori Shirakawa and Lukas Broers and Han Xu and Miwako Tsuji and Mitsuhisa Sato and Seiji Yunoki and Ryo Wakizaka and Yukio Kawashima and Jun Doi and Toshinari Itoko and Hiroshi Horii and Thaddeus Pellegrini and Javier {Robledo Moreno} and Kevin J. Sung and Ella Fejer and Robert Walkup and Seetharami Seelam and Mario Motta},
      year={2026},
      eprint={2605.01138},
      archivePrefix={arXiv},
      primaryClass={quant-ph},
      url={https://arxiv.org/abs/2605.01138}, 
}

@article{knill2008,
  author  = {Knill, Emanuel and Leibfried, Dietrich and Reichle, Rolf and Britton, Joseph and Blakestad, R. Brad and Jost, John D. and Langer, Christopher and Ozeri, Roee and Seidelin, Signe and Wineland, David J.},
  title   = {Randomized Benchmarking of Quantum Gates},
  journal = {Physical Review A},
  volume  = {77},
  pages   = {012307},
  year    = {2008},
  doi     = {10.1103/PhysRevA.77.012307}
}

@article{magesan2012,
  author  = {Magesan, Easwar and Gambetta, Jay M. and Emerson, Joseph},
  title   = {Characterizing Quantum Gates via Randomized Benchmarking},
  journal = {Physical Review A},
  volume  = {85},
  pages   = {042311},
  year    = {2012},
  doi     = {10.1103/PhysRevA.85.042311}
}

@article{henning2006,
  author  = {Henning, John L.},
  title   = {{SPEC CPU2006} Benchmark Descriptions},
  journal = {ACM SIGARCH Computer Architecture News},
  volume  = {34},
  number  = {4},
  pages   = {1--17},
  year    = {2006},
  doi     = {10.1145/1186736.1186737}
}

@inproceedings{amdahl1967,
  author    = {Amdahl, Gene M.},
  title     = {Validity of the Single Processor Approach to Achieving Large Scale Computing Capabilities},
  booktitle = {Proceedings of the Spring Joint Computer Conference (AFIPS)},
  pages     = {483--485},
  year      = {1967},
  doi       = {10.1145/1465482.1465560}
}

@article{kim2023,
  author  = {Kim, Youngseok and Eddins, Andrew and Anand, Sajant and Wei, Ken Xuan and {van den Berg}, Ewout and Rosenblatt, Sami and Nayfeh, Hasan and Wu, Yantao and Zaletel, Michael and Temme, Kristan and Kandala, Abhinav},
  title   = {Evidence for the utility of quantum computing before fault tolerance},
  journal = {Nature},
  volume  = {618},
  pages   = {500--505},
  year    = {2023},
  doi     = {10.1038/s41586-023-06096-3}
}

@article{lubinski2023,
   title={Application-Oriented Performance Benchmarks for Quantum Computing},
   volume={4},
   ISSN={2689-1808},
   url={http://dx.doi.org/10.1109/TQE.2023.3253761},
   DOI={10.1109/tqe.2023.3253761},
   journal={IEEE Transactions on Quantum Engineering},
   publisher={Institute of Electrical and Electronics Engineers (IEEE)},
   author={Lubinski, Thomas and Johri, Sonika and Varosy, Paul and Coleman, Jeremiah and Zhao, Luning and Necaise, Jason and Baldwin, Charles H. and Mayer, Karl and Proctor, Timothy},
   year={2023},
   pages={1--32} }

@misc{cosentino2026,
      title={Metriq: A Collaborative Platform for Benchmarking Quantum Computers}, 
      author={Alessandro Cosentino and Changhao Li and Vincent Russo and Bradley A. Chase and Tom Lubinski and Siyuan Niu and Neer Patel and Nathan Shammah and William J. Zeng},
      year={2026},
      eprint={2603.08680},
      archivePrefix={arXiv},
      primaryClass={quant-ph},
      url={https://arxiv.org/abs/2603.08680}, 
}

@article{vandenberg2023,
  author  = {{van den Berg}, Ewout and Bravyi, Sergey and Gambetta, Jay M. and Jurcevic, Petar and Maslov, Dmitri and Temme, Kristan},
  title   = {Single-Shot Error Mitigation by Coherent {Pauli} Checks},
  journal = {Physical Review Research},
  volume  = {5},
  pages   = {033193},
  year    = {2023},
  doi     = {10.1103/PhysRevResearch.5.033193},
  eprint  = {2212.03937},
  archivePrefix = {arXiv},
  primaryClass  = {quant-ph}
}

@article{gonzales2023,
   title={Quantum error mitigation by Pauli check sandwiching},
   volume={13},
   ISSN={2045-2322},
   url={http://dx.doi.org/10.1038/s41598-023-28109-x},
   DOI={10.1038/s41598-023-28109-x},
   number={1},
   pages={2122},
   journal={Scientific Reports},
   publisher={Springer Science and Business Media LLC},
   author={Gonzales, Alvin and Shaydulin, Ruslan and Saleem, Zain H. and Suchara, Martin},
   year={2023},
   month=Feb }

@misc{martiel2025,
      title={Low-overhead error detection with spacetime codes}, 
      author={Simon Martiel and Ali Javadi-Abhari},
      year={2025},
      eprint={2504.15725},
      archivePrefix={arXiv},
      primaryClass={quant-ph},
      url={https://arxiv.org/abs/2504.15725}, 
}

@misc{merkel2025,
      title={When Clifford benchmarks are sufficient; estimating application performance with scalable proxy circuits}, 
      author={Seth Merkel and Timothy Proctor and Samuele Ferracin and Jordan Hines and Samantha Barron and Luke C. G. Govia and David McKay},
      year={2025},
      eprint={2503.05943},
      archivePrefix={arXiv},
      primaryClass={quant-ph},
      url={https://arxiv.org/abs/2503.05943}, 
}

@article{nation2025,
   title={Benchmarking the performance of quantum computing software for quantum circuit creation, manipulation and compilation},
   volume={5},
   ISSN={2662-8457},
   url={http://dx.doi.org/10.1038/s43588-025-00792-y},
   DOI={10.1038/s43588-025-00792-y},
   number={5},
   journal={Nature Computational Science},
   publisher={Springer Science and Business Media LLC},
   author={Nation, Paul D. and Saki, Abdullah Ash and Brandhofer, Sebastian and Bello, Luciano and Garion, Shelly and Treinish, Matthew and Javadi-Abhari, Ali},
   year={2025},
   month=Apr, pages={427--435} }

@misc{ibmhardware2026,
  author       = {David McKay and Scott Crowder and Jerry Chow and Oliver Dial and Robert Davis and Catherine Dundon},
  title        = {How to measure the performance of a quantum computer},
  year         = {2026},
  howpublished = {IBM Quantum Blog},
  note         = {\url{https://www.ibm.com/quantum/blog/hardware-metrics-2026}}
}

\end{document}